\documentclass[twocolumn]{webofc}

\usepackage[varg]{txfonts}   
\usepackage{hyperref}
\usepackage{url}
\usepackage{color}
\usepackage{subfigure}

\newcommand\br{{\bf r}}
\newcommand\bR{{\bf R}}
\newcommand\brho{{\boldsymbol \rho}}
\newcommand\bO{{\bf O}}

\newcommand{\beq}{\begin{equation}}
\newcommand{\eeq}{\end{equation}}

\hypersetup{colorlinks=true,citecolor=blue,urlcolor=blue,linkcolor=blue}
\begin{document}
\title{Self-organization, detailed balance, and stress-structure correlations in 2D granular dynamics}
%
%

\author{\firstname{Raphael} \lastname{Blumenfeld}\inst{1}\fnsep\thanks{\email{rbb11@cam.ac.uk}} 
\and \firstname{Takashi} \lastname{Matsushima}\inst{2}
\and \firstname{Jie} \lastname{Zhang}\inst{3}
}

\institute{Gonville \& Caius College, University of Cambridge, Trinity Street, Cambridge CB2 1TA, UK
\and Faculty of Engineering, Information and Systems, University of Tsukuba, Ibaraki 305-8573, Japan
\and School of Physics and Astronomy, Shanghai Jiao Tong University, Shanghai 200240, China }

\abstract{We argue that a suite of recent experimental and numerical observations are signatures of cooperative stress-structure self-organisation (SO) in granular dynamics. These observations include: a) detail-insensitive collapses of certain quantities; b) correlations between stress and structure and evidence of entropy-stability competition in settled packings, which cast doubt on most linear stress theories of granular materials; c) detailed balanced steady states, which seem contradictory to the common belief that only systems in thermal equilibrium satisfy detailed balance, but are not, as we explain. 
We then propose a new statistical mechanical formulation that takes into account the cooperative SO.} 

\maketitle

\section{Introduction}
The macroscale behavior of granular media is sensitive to grain-scale characteristics, severely limiting first-principles continuum models. Consequently, existing models resort to phenomenological and empirical constitutive properties. In particular, all stress models rely on assumed constitutive properties that are independent of the stress state. This assumption ignores effects of cooperative structure-stress SO~\cite{tordesillas2009modeling,Bletal2015,matsushima2014universal,matsushima2017fundamental,Sunetal20}. Stress-dependent constitutive properties imply that the stress field equations must be nonlinear, casting doubt on most existing models~\cite{MaBl21}. 
Work on upscaling stress-structure relations to the continuum were based on force chains \cite{SeeligWulff1946, oda1985stress, cates1998jamming}, fabric tensors \cite{Rothenburg1989fabric, BaBl2002, Bl04, BlEd06, Nguyen2015fabric}, and other microstructural characteristics~\cite{tordesillas2010force, Donev2007-lc,Bl20}, but reliable upscaling requires understanding of the cooperative SO at the grain-scale. 
Relevant to such an understanding are recent observations of collapses of various distributions for a range of parameters and protocols, emergence of detailed balanced steady states~\cite{Waetal20,Sunetal21}, discoveries of entropy-stability competition~\cite{Sunetal20}, and stress-structure correlations~\cite{MaBl21,Jietal22}. 
We describe these briefly here, argue that they necessitate new modeling approaches, and propose a new statistical mechanical formulation to account for the stress-structure correlations. 

\section{Detailed-balanced steady states (DBSS) in cell dynamics}
\subsection{Theory}
Granular structures evolve by making and breaking contacts. These merge and split cells, which are the elementary structural elements of the stress-carrying 'backbone'. In 2D, a cell order has been defined as the number of grains in contacts surrounding it~\cite{Waetal20}. The cell order, cell stress, and shape distributions were found to provide insight into the structural SO~\cite{tordesillas2009modeling,Bletal2015,matsushima2014universal,matsushima2017fundamental}.  \\
We focus here on the backbone and ignore rattlers. Including them in the evolution equations is possible, but this complicates the analysis without providing further insight~\cite{Waetal20}. Moreover, at steady state of quasi-static dynamics, the mean number of particles in the backbone is constant and measuring backbone characteristics at  uncorrelated moments provides good statistics. Indeed, it was shown in~\cite{Sunetal20} that neglecting rattlers does not affect conclusions about the steady-state backbone characteristics. We also note in passing that rattlers are irrelevant when studying systems subject to gravity. 
To simplify notations below, we index cells of order $k$ as $c_{k-2}$. Thus, when the contact separating cells $c_n$ and $c_m$ breaks, they merge into $c_{n+m}$ and vice versa when a contact is made, as illustrated in Fig.~\ref{TransitionDBSS}. The CID dynamics are
\begin{align}
   \dot{Q}_k = 
       & \frac{1}{2} \sum_{n=1}^{k-1} \eta_{n,k-n} (1+\delta_{n,k-n}) \notag \\
       & - \sum_{m=k+1}^{\infty} \eta_{k,m-k} (1+\delta_{k,m-k})
       + Q_k \sum_{\substack{i,j=1\\i\leq j}}^{\infty} \eta_{i,j} \ .
       \label{CIDEvolution}
\end{align}
$\eta_{i,j}\equiv p_{i,j} Q_iQ_j - q_{i,j}Q_{i+j}$ measures the direction of the process $c_i+c_j\leftrightharpoons c_{i+j}$, $p_{i,j}$ is the merging rate of $c_i$ and $c_j$, and $q_{i,j}$ the reverse splitting rate of $c_{i+j}$. The last term on the r.h.s. of (\ref{CIDEvolution}) accounts for the change in the overall number of cells during the dynamics~\cite{Waetal20}. 
The $\eta_{i,j}$ processes are equivalent to chemical reactions, with same-index cells mapping into same-substance molecules. 

\subsection{Experimental test of DBSS}
Using an 'infinite shear' stadium device~\cite{Mietal13}, we tested the validity of eqs. (\ref{CIDEvolution}) in an experiment, described in detail in~\cite{Sunetal20,Sunetal21}. We summarise it briefly next for completeness. The device consisted of two stainless steel sprockets connected by a corrugated rubber belt, which confined by a low confining pressure a packing of flat particles on a solid surface. A stepping motor drove the sprockets cyclically, shearing the particles by the belt, and the dynamics in the central region was recorded by a video camera. The particles were sheared cyclically continuously at strains that ranged from $3\%$ to $10\%$, which exceeded the yielding strain of $\sim3\%$. 
To investigate friction effects, several particle types were used: gear-shaped nylon particles ($\mu \rightarrow \infty$), combinations of stainless-steel rings and gear particles, photoelastic disks ($\mu \approx 0.7$), ABS plastic rings ($\mu \approx 0.32$), stainless-steel rings ($\mu \approx 0.3$), and Teflon-coated photoelastic disks ($\mu \approx 0.15$). Crystallization was minimised by using $50-50$ binary mixtures of sizes $1.0$cm and $1.4$cm ($1.6$cm and $1.84$cm pitch diameter for gear particles). Each system contained approximately 2,000 particles. \\
Particles were deposited randomly on the surface and a cyclic shear was applied a steady state was reached, typically within 100 cycles. The process was paused at maximum negative strain and snapshots of the structure were taken. Steady state was determined by noting that all measurable quantities fluctuated around a constant mean. The packing fractions were: $0.74\pm0.01$ for gears, $0.81\pm0.01$ for photoelastic particles, and $0.83\pm0.01$ for Teflon-coated photoelastic, stainless steel, and plastic particles - consistent with the expectation that higher friction reduces packing fraction under constant pressure.

\subsection{Observations of DBSS}
\begin{figure}[h]
\begin{center}
\includegraphics[width=0.45\textwidth]{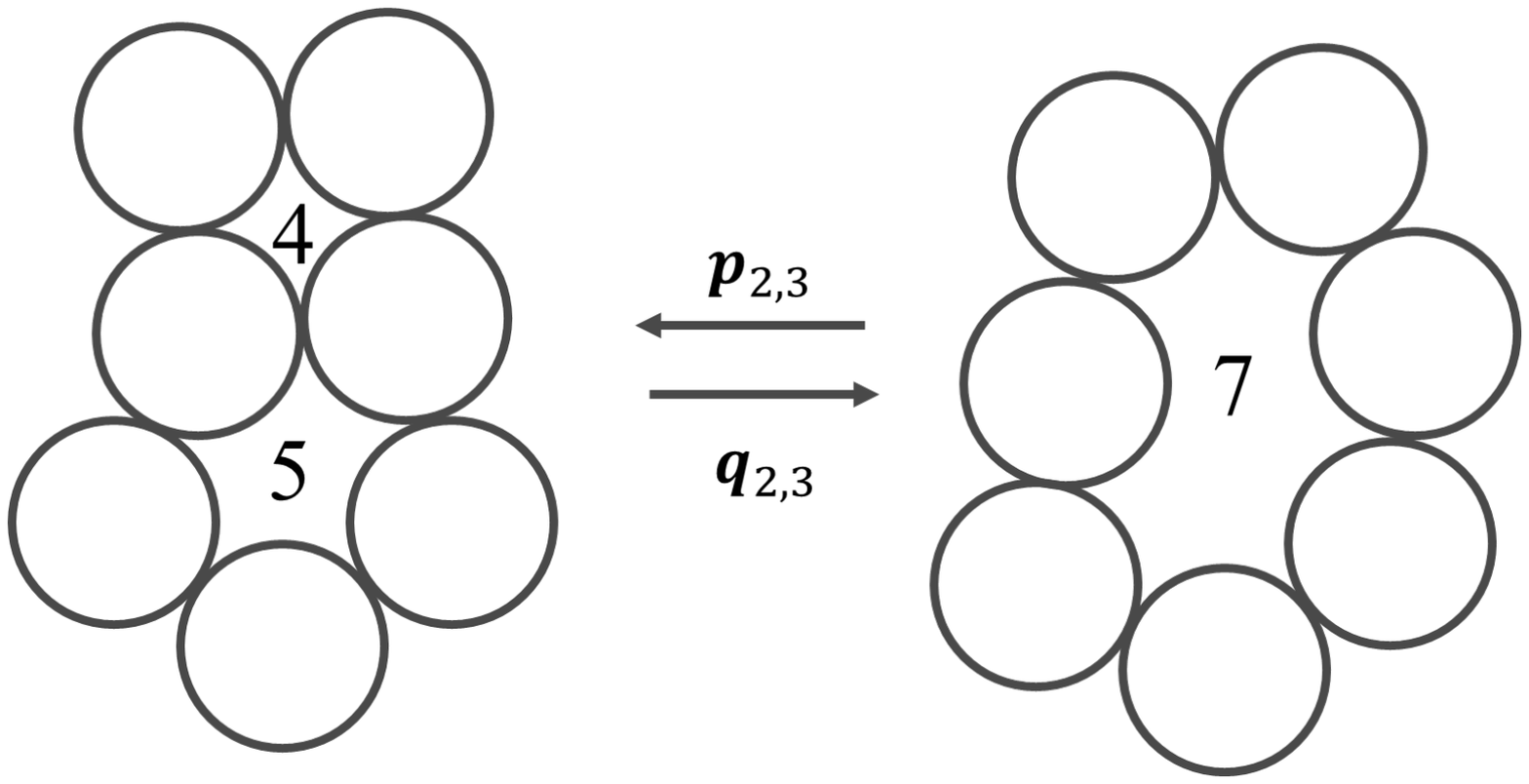}
\includegraphics[width=0.45\textwidth]{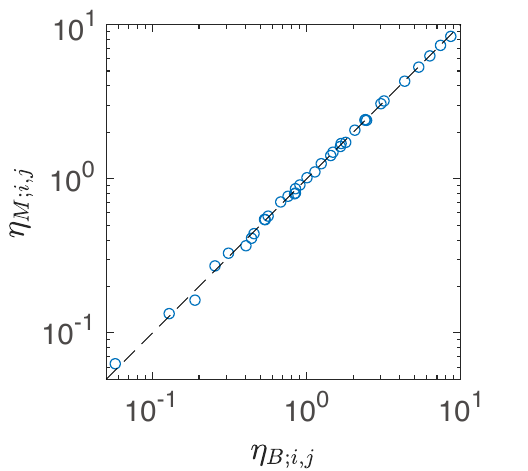}
\caption{Top: an illustration of merging and breaking transitions involving cells of order $4$ and $5$, whose indices are, respectively, $k=4-2=2$ and $k=5-2=3$. The related processes are $c_2+c_3\leftrightharpoons c_{5 }$.  $p_{2,3}$ is the rate of merging of cells of indices $k=2$ and $k=3$ int a cell of index $k=5$, while $q_{2,3}$ is the rate of splitting of the latter into the former two.
Bottom: the merging, $\eta_{M;i,j}$, and breaking, $\eta_{B;i,j}$, rates for all orders up to 10 tend detailed balance as sample size increased (the axes denote orders).}
\label{TransitionDBSS}
\end{center}
\end{figure}

The experiments and simulation revealed a surprise: as sample sizes increased, the dynamics converged to DBSS - each and every process became balanced at steady-state, $\eta_{i,j}=0$ $\forall i,j$, statistically significant up to order 10 (FIg.~\ref{TransitionDBSS}).  The same qualitative results were obtained in a similar experiment and numerical simulation, in which the particles confined to a circular apparatus~\cite{,Mayetal24}.\\
The observation of DBSS appears to defy the common belief that only thermally equilibrated systems satisfy detailed balance, while non-equilibrium steady states must include process cycles, e.g., $A\to B\to C\to A$~\cite{Klein55}. But there is no contradiction. Firstly, Klein's proof applied only to thermal systems, while this is an a-thermal one. Secondly, unlike in thermal equilibrium, where detailed balance arises spontaneously on experimental timescales, our observed DBSS are emergent and result from SO of the spatial distribution of the cell indices~\cite{Sunetal21,Mayetal24}, which rearranges to enable detailed balance~\cite{Mayetal24}. In parallel, the rates $p_{i,j}$ and $q_{i,j}$ self-organise toward values that maintain DBSS~\cite{Sunetal21}. 

\section{SO in cell stress and structure}
With the cell being a key element of the structure, it is important to define both cell stress and cell structure. To this end, we first review briefly the quadron-based quantification of the structure around cells~\cite{BaBl2002,Bl04,BlEd06}. Regarding inter-granular contacts as points (line and double contacts can also be represented by points~\cite{Bletal2015,DContact05}), this is done as follows. We determine the centroid of the $z_g$ contacts around any grain $g$, $\brho_g$, and connect them by vectors $\br_{c,g}$ ($c=1,2,...,z_g$) circulating clockwise. The number of cells surrounding $g$ is also $z_g$ and we identify the centroid of the contacts surrounding each of these cells, $\bO_c$. Each vector $\br_{c,g}$ is shared between grain $g$ and one of its neighbour cells, $c$. Next, extend a vector from $\brho_g$ to $\bO_c$, $\bR_{c,g}=\bO_c-\brho_g$. The vectors $\br_{c,g}$ and $\bR_{c,g}$ form the diagonals of a rectangle, called quadron, whose area, $A_{c,g}=|\br_{c,g}\times\br_{c,g}|/2$, is shared between cell $c$ and grain $g$. Repeating this procedure for each grain, the quadrons tessellate the entire planar system without gaps or overlaps~\cite{QOverlap}. The grain and cell areas are defined as 
\begin{equation}
A_{g} = \sum_{c=1}^{z_g}A_{c,g}   \quad ; \quad A_{c} = \sum_{g=1}^{z_c} A_{c,g}
\label{Ac}
\end{equation}
with the rightmost sum over the $z_c$ grains surrounding $c$. 

\subsection{Definition of cell stress}
To define now cell stresses, we start from the standardly-defined stress on grain $g$, 
\begin{equation}
\sigma_{g} = \frac{1}{A_g}\sum_{g'} f_{g',g}\otimes \br_{g',g} \ .
\label{Sigg}
\end{equation}                                         
The cell stress is a weighted sum over the grains surrounding it~\cite{MaBl21,Kretal25},
\begin{equation}
\sigma_{c} = \frac{1}{A_c}\sum_{g=1}^{z_c} A_g\sigma_g \ .
\label{Sigc}
\end{equation}
We also define a measure of cell stability,
\beq
h_c\equiv\frac{\sigma_{1,c}-\sigma_{2,c}}{\sigma_{1,c}+\sigma_{2,c}} \ ,
\label{h}
\eeq 
where $\sigma_{1,c}$ and $\sigma_{2,c}\leq\sigma_{1,c}$ are the cell principal stresses. \\
An analysis of the conditional distributions of $h_c$, given the cell index, $n$, when each is scaled by its mean, $\bar{h}_c(n)$, revealed that all the conditional distributions collapse onto one curve that has a Weibull form independent of $n$
\begin{equation}
P\left[x\equiv\frac{h_c(n)}{\bar{h}_c(n)}\mid n\right] = K x^{\alpha-1} e^{-\left(x/\lambda\right)^\alpha} \ .
\label{Weibull}
\end{equation}
$\alpha$ and $\lambda$ are constants and $K(\alpha,\lambda)$ is a normalisation factor. 
A model for the self-organised emergence of this particular form has been proposed recently~\cite{Jietal22}.

\subsection{Cooperative stress-shape SO and entropy-stability competition}
Another signature of stress-structure SO emerge from considering cell configurational entropy. Mechanical stability constrains entropy by limiting the number of realisable configurations. Using a normalised mean area as an indicator of same-index cells in monodisperse packings, it was demonstrated that the entropy first increases with the index, but then decreases~\cite{matsushima2017fundamental,Sunetal20}. It was also shown that in the absence of any stress field the entropy increases monotonically~\cite{matsushima2017fundamental}. \\
Arguably, the clearest signature of the cooperative stress-structure SO comes from investigations into the effects of stress on individual cell configurations. Approximating each cell as an ellipse of axes $a_c$, $b_c$, and $\theta_c$, it was found that each cell's ellipse aligns well with the orientation of its principal stress, $\theta^\sigma_c$, with the deviation, $\Delta\theta\equiv\theta^\sigma_c-\theta_c$, distributed narrowly around $0$ for all cell indices~\cite{MaBl21,Jietal22}. This strong correlation between cells' shapes and their stresses is a direct result of a cooperative SO. Cells of higher index are more strongly correlated because they are more easily reorientated and have a narrower range of stability to torques during dynamics.

\section{Statistical mechanics of stress-structure interactions}
The involvement of cells in all the above signatures of SO suggests that they are a key structural element and it is then appropriate to focus on them to further understanding of granular systems. To this end, we propose the fiollowing statistical mechanical model that includes the interactions between cell shapes and stresses. Using the ellipse approximation, a cell has six degrees of freedom: $a_c$, $b_c$, $\theta_c$, $\sigma_1$, $\sigma_2$, and $\theta_\sigma$. The correlations and the stability-entropy competition can be reproduced by coupling terms.
A possible grand-canonical partition function is
\begin{align}
Z=\prod\limits_{k=1}^\infty \left[\left(\sum_{c\in k}^{n_c(k)} e^{-F_c-G_c}\right) e^{-\mu n_c(k)}\right] \ .
\label{Z}
\end{align}
The sum is over all the $n_c(k)$ $k$-cells, $k$ running over the indices, 
\begin{align}
F_c = \alpha_1 a_c + \alpha_2 b_c + \alpha_3\theta_c + \beta_1 \sigma_{1,c} + \beta_2 \sigma_{2,c} + \beta_3\theta_\sigma
\label{Fc}
\end{align}
controls the entropy of the degrees of freedom of $k$-cells, 
\begin{align}
G_c = J_1 \frac{a_c - b_c}{a_c + b_c} + J_2 \frac{\sigma_{1,c} - \sigma_{2,c}}{\sigma_{1,c} + \sigma_{2,c}} + J_3 \left(\theta_c - \theta_\sigma\right)^2
\label{Fc}
\end{align}
comprises the couplings between them, informed by the above observations. $\alpha_i$, $\beta_i$, and $\mu$ are the usual Lagrange multipliers, and the $J_i$ are coupling constants. 
The sum over the $k$-cells can be converted into integrals over `densities of states', some of which are known already: $\theta_c$ is isotropic, $\theta_\sigma$ in simple shear is normally distributed around $\pi/4$, $\theta_c - \theta_\sigma$ is normally distributed around zero, and $h$ has a Weibull distribution.

\section{Conclusion, discussion, and outlook}
To conclude, the observed detailed balance, collapsed  distributions, entropy-stability competition, and stress-structure correlations are all signatures of the cooperative SO in granular systems. 
One significant implication of these observations and result is that the dependence of the local structure on the local stress makes the local constitutive properties in any stress field theory stress-dependent. This means that the stress field equations to model granular materials must be non-linear, undermining most existing linear stress theories, including critical state theory~\cite{ScWr68}. \\
Another important aspect of the observations reported here is the emergence of DBSS. The detailed balance principle led to enormous progress in traditional physics models and it is likely that applying it in granular science could lead to similar progress. This potential is exciting and deserves further exploration.  \\
The results further point to a new way to extend the idea of a statistical mechanical modelling of granular materials~\cite{EdOa89,BlEd09,Bletal16} by including the cooperativity of stress and structure. We have proposed here a grand-canonical partition function, in which cells are virtual quasi-particles that can appear and disappear and cell stresses and shapes interact. The work on this model has started and results will be reported in due course.\\
Finally, some of the SO signatures discussed here are straightforward to look for in three dimensions. These are the definition of cell stresses and approximation of cell shapes as ellipsoids, extensions that would provide the data to determine stress-shape correlations in cells, stability-entropy competition, and potential collapses of distributions as in two dimensions. Work in this direction is currently under way. Determining detailed balance is also possible, but requires much more work because a new (and cumbersome) classification of cells is needed, e.g., by the numbers of facets and facet indices. Such a search is probably achievable with computational assistance.

\end{document}